\documentstyle[psfig,epsf,conf-X,10pt]{article}
\begin{document} 
\small
\heading{%
%
Implications of Abundance Gradients in Intracluster Gas
}
\par\medskip\noindent
\author{%
Raymond E. White III$^{1,2}$, Renato A. Dupke$^{3}$
}
\address{%
Laboratory for High Energy Astrophysics, NASA GSFC, Greenbelt, MD 20771
}
\address{%
Department of Physics \& Astronomy, University of Alabama,
Tuscaloosa, AL 35487-0324; white@merkin.astr.ua.edu
}
\address{%
Department of Astronomy, University of Michigan, Ann Arbor, MI 48109-1090;
rdupke@umich.edu
}

\begin{abstract}
Analysis of spatially resolved {\sl ASCA} spectra of the intracluster gas
in Abell 496 confirms that metal abundances increase toward the center.
We also find spatial gradients in several abundance $ratios$, indicating that
the fraction of iron from SN Ia increases toward the cluster center and that
the dominant metal enrichment mechanism near the center
must be different than in the outer parts.
\end{abstract}
\section{Introduction}
While the metals observed in intracluster gas clearly originate from stars, 
how the metals got from stars into the intracluster gas remains controversial. 
The two global metal enrichment mechanisms most commonly considered are 
supernovae-driven protogalactic winds from early-type galaxies\cite{LD} and 
ram-pressure stripping of gas from cluster galaxies\cite{GG}. 
At the centers of cD clusters, accumulated stellar mass loss from the cDs
may also contribute to the observed metal distribution.
In principle, the imprints of these enrichment mechanisms can be
distinguished by the chemical mix and spatial distribution of heavy elements
in intracluster gas.
Protogalactic winds, powered by Type II supernovae from early generations of 
short-lived, massive stars, may distribute metals throughout clusters.
Ram-pressure stripping would be most effective near cluster centers and
should deposit gas from galaxy atmospheres with considerable supplemental 
enrichment from Type Ia supernovae, since stripping is a more secular, 
ongoing process; SN Ia have longer-lived progenitors 
(accreting white dwarfs) and different elemental yields than SN II.
Accumulated stellar mass loss in central cDs should have somewhat higher
abundances than gas stripped from other early-type galaxies (but similar
abundance ratios) and may have a different spatial profile than stripped gas.

Unfortunately, residual uncertainty in the theoretical elemental yields from 
SN II have allowed different interpretations of recent {\sl ASCA} spectroscopy of 
intracluster gas. The yield models adopted by some investigators\cite{ML}\cite{MLA}
imply that global intracluster metal abundances are consistent with  
SN II ejecta, supporting the protogalactic wind enrichment scenario.
However, we and others, using different theoretical yield models for SN II,  
find that as much as 50\% of intracluster iron comes from 
SN Ia\cite{DW1}\cite{DW2}\cite{IA}\cite{NS}. 
Somewhat more than half of the global iron comes from SN II, which
can be readily attributed to protogalactic wind enrichment.
However, the presence of large quantities of iron from SN Ia throughout 
clusters is problematic: is ram pressure stripping so effective that it 
contaminates the outer parts of clusters as much as the central regions?

Since the detailed spatial distribution of elements in intracluster gas
may offer clues about the dominant metal enrichment mechanism(s),
we analyzed {\sl ASCA} observations of Abell 496: analysis of previous 
{\sl Ginga} and {\sl Einstein} X-ray satellite data indicated 
that it has centrally enhanced abundances\cite{WD}.

\begin{figure}
\includegraphics{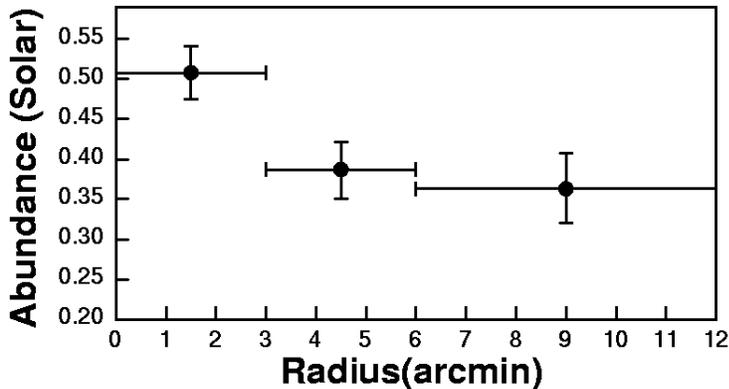}
\centerline{\null}
\centerline{\null}
\vskip1.8in
\caption[]{Elemental abundance distribution (driven by iron) in Abell 496.}
\label{fig:agrad}
\end{figure}

\section{Analysis of Abell 496}

We jointly fitted isothermal emissivity models to spatially resolved
spectra of Abell 496 from all four {\sl ASCA} instruments.
Tying individual elemental abundances together in these fits, we
find the metal abundance increases from 0.36 solar in the outer
$3-12'$ region to 0.51 solar within $3'$ (see Fig.~\ref{fig:agrad}).
Allowing the abundances of individual elements to vary independently,
we find that iron, nickel and sulfur abundances are centrally 
enhanced.  Our results are the same when we include a cooling flow
spectral component for the emission from the central region.

We also found significant gradients in several abundance $ratios$:
Si/S, Si/Ni and S/Fe.
Having gradients in abundance ratios implies that the
proportion of SN Ia/II ejecta is changing spatially and that
the dominant metal enrichment mechanism(s) near the cluster center
must be different than in the outer parts.

We compared an ensemble of observed abundance ratios to theoretical
expectations of yields from SN Ia\cite{NI}\cite{NTY} and SN II\cite{NH} 
in order to estimate the relative proportion of SN Ia/II ejecta in Abell 496.
An ensemble of ratios is required, since there are large theoretical 
uncertainties in the yields of individual elements.

\begin{figure}
\includegraphics{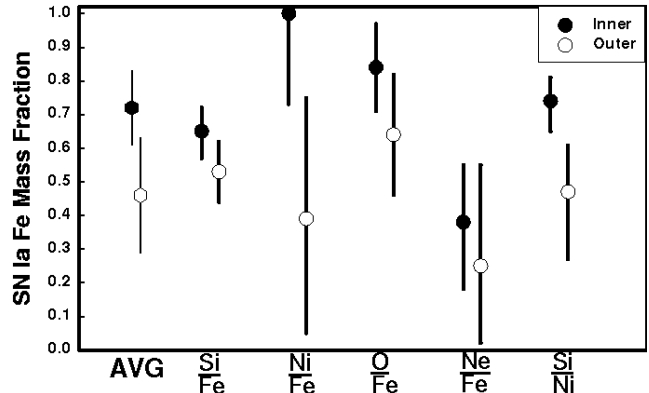}
\centerline{\null}
\centerline{\null}
\vskip1.8in
\caption[]{Iron mass fraction from SN Ia in inner ($0-2'$) and outer 
regions ($3-12'$) of Abell 496, from an ensemble of elemental
abundance ratios.
}
\label{fig:snfrac}
\end{figure}

Fig.~\ref{fig:snfrac} illustrates various estimates of the iron mass fraction 
due to SN Ia in inner ($0-2'$; filled circles) and outer ($3-12'$; empty
circles) regions of the cluster.
The ensemble of the five best-determined abundance ratios collectively
and individually indicate that the proportion of SN Ia ejecta increases
toward the center.  
The average of the five individual estimates of the iron mass fraction
from SN Ia is also indicated in Fig.~\ref{fig:snfrac}: 
the proportion of ejecta from
SN Ia is $\sim46$\% in the outer parts, rising to $\sim72$\% within $2'$.
The central increase in the proportion of SN Ia ejecta is such that the 
central iron abundance enhancement can be attributed entirely to SN Ia ejecta.

				\section{
Enrichment Mechanisms
				}

The central metal abundance enhancements in Abell 496 are not likely to be 
caused by ram pressure stripping of gas from cluster galaxies.
The gaseous abundances measured in most early-type galaxies by
{\sl ASCA}\cite{LM}\cite{ME} and {\sl ROSAT}\cite{DW} are 0.2-0.4 solar, 
significantly less than the 0.5-0.6 solar abundance observed at
the cluster center.
Only the most luminous ellipticals, which also tend to be at the
centers of galaxy clusters or groups, are observed to have gaseous 
abundances of 0.5-1 solar.
If ram pressure stripping is effective in the cluster, it would act
to $dilute$ the central abundance enhancement.
If ram pressure stripping is not the primary source of metals
near the cluster center, where it should be most effective, 
it is an even less likely source of metals
in the outer parts of the cluster.

The central abundance enhancement is also unlikely to be due to
the secular accumulation of stellar mass loss in the cD.
If it were, then other giant ellipticals which have not been
stripped should have comparable ratios of iron mass to optical
luminosity.  NGC 4636 is among the most X-ray luminous ellipticals 
and may be at the center of its own small group.  However, NGC 4636
has a $\sim$10-20 times smaller iron mass to light ratio in its vicinity
compared to the cD of Abell 496.

We propose instead that the bulk of intracluster gas is contaminated 
by two phases of winds from early-type galaxies:
an initial SN II-driven protogalactic wind phase, followed by a secondary,
less vigorous SN Ia-driven wind phase, contaminating the bulk of
intracluster gas with comparable masses of iron.
The secondary SN Ia-driven winds would 
be $\sim$ 10 times less energetic than the initial SN II-driven
protogalactic winds, since SN Ia inject $\sim$10 times less energy per unit 
iron mass than SN II and the observations indicate that comparable amounts
of iron came from SN Ia and SN II.
Less vigorous secondary SN Ia-driven winds would allow SN Ia-enriched
material to escape most galaxies, but not clusters.
However, the secondary SN Ia-driven wind from a central dominant galaxy
may be smothered, due to the galaxy's location
at the bottom of the cluster's gravitational 
potential and in the midst of the highest ambient intracluster gas density.
Such a smothered wind could generate the metal abundance enhancement seen
at the center of Abell 496, which has the chemical signature of SN Ia ejecta.

\acknowledgements{This work was supported by NASA grant NAG 5-2574
and a National Research Council Senior Research Associateship at NASA GSFC.
}

\vfill
\end{document}